\documentclass[superscriptaddress,showpacs,twocolumn,prd]{revtex4}
\usepackage{latexsym}
\usepackage{graphicx}

\newcommand{\nc}{\newcommand}

\nc{\be}[1]{\begin{equation}\mbox{$\label{#1}$}}
\nc{\bea}[1]{\begin{eqnarray} \mbox{$\label{#1}$}}
\nc{\Section}[2]{\section{#2}\label{#1}}
\nc{\Bibitem}[1]{\bibitem{#1}}
\nc{\Label}[1]{\label{#1}}

\nc{\eea}{\end{eqnarray}}
\nc{\ee}{\end{equation}}

\nc{\bdm}{\begin{displaymath}}
\nc{\edm}{\end{displaymath}}
\nc{\dpsty}{\displaystyle}
\nc{\bc}{\begin{center}}
\nc{\ec}{\end{center}}
\nc{\ba}{\begin{array}}
\nc{\ea}{\end{array}}
\nc{\bab}{\begin{abstract}}
\nc{\eab}{\end{abstract}}
\nc{\btab}{\begin{tabular}}
\nc{\etab}{\end{tabular}}
\nc{\bit}{\begin{itemize}}
\nc{\eit}{\end{itemize}}
\nc{\ben}{\begin{enumerate}}
\nc{\een}{\end{enumerate}}
\nc{\bfig}{\begin{figure}}
\nc{\efig}{\end{figure}}

\nc{\arreq}{&\!=\!&}
\nc{\arrmi}{&\!-\!&}
\nc{\arrpl}{&\!+\!&}
\nc{\arrap}{&\!\!\!\approx\!\!\!&}
\nc{\non}{\nonumber}
\nc{\align}{\!\!\!\!\!\!\!\!&&}

\def\lsim{\; \raise0.3ex\hbox{$<$\kern-0.75em
      \raise-1.1ex\hbox{$\sim$}}\; }
\def\gsim{\; \raise0.3ex\hbox{$>$\kern-0.75em
      \raise-1.1ex\hbox{$\sim$}}\; }

\nc{\DOT}{\hspace{-0.08in}{\bf .}\hspace{0.1in}}
\nc{\Laada}{\hbox {$\sqcap$ \kern -1em $\sqcup$}}
\nc\loota{{\scriptstyle\sqcap\kern-0.55em\hbox{$\scriptstyle\sqcup$}}}
\nc\Loota{{\sqcap\kern-0.65em\hbox{$\sqcup$}}}
\nc\laada{\Loota}
\nc{\qed}{\hskip 3em \hbox{\BOX} \vskip 2ex}

\nc{\real}{{\rm I \! R}}
\nc{\Z}{{\sf Z \!\!\! Z}}
\nc{\complex}{{\rm C\!\!\! {\sf I}\,\,}}
\def\bigid{\leavevmode\hbox{\small1\kern-3.8pt\normalsize1}}
\def\id{\leavevmode\hbox{\small1\kern-3.3pt\normalsize1}}
\nc{\slask}{\!\!\!/}
\nc{\bis}{{\prime\prime}}
\nc{\pa}{\partial}
\nc{\na}{\nabla}
\nc{\ra}{\rangle}
\nc{\la}{\langle}
\nc{\goto}{\rightarrow}
\nc{\swap}{\leftrightarrow}

\nc{\EE}[1]{ \mbox{$\cdot10^{#1}$} }
\nc{\abs}[1]{\left|#1\right|}
\nc{\at}[2]{\left.#1\right|_{#2}}
\nc{\norm}[1]{\|#1\|}
\nc{\abscut}[2]{\Abs{#1}_{\scriptscriptstyle#2}}
\nc{\vek}[1]{{\rm\bf #1}}
\nc{\integral}[2]{\int\limits_{#1}^{#2}}
\nc{\inv}[1]{\frac{1}{#1}}
\nc{\dd}[2]{{{\partial #1}\over{\partial #2}}}
\nc{\ddd}[2]{{{{\partial}^2 #1}\over{\partial {#2}^2}}}
\nc{\dddd}[3]{{{{\partial}^2 #1}\over
    {\partial #2 \partial #3}}}
\nc{\dder}[2]{{{d #1}\over{d #2}}}
\nc{\ddder}[2]{{{d^2 #1}\over{d {#2}^2}}}
\nc{\dddder}[3]{{d^2 #1}\over
    {d #2 d #3}}
\nc{\dx}[1]{d\,^{#1}x}
\nc{\dy}[1]{d\,^{#1}y}
\nc{\dz}[1]{d\,^{#1}z}
\nc{\dl}[1]{\frac{d\,^{#1}l}{(2\pi)^{#1}}}
\nc{\dk}[1]{\frac{d\,^{#1}k}{(2\pi)^{#1}}}
\nc{\dq}[1]{\frac{d\,^{#1}q}{(2\pi)^{#1}}}

\nc{\bfT}{{\bf T }}

\nc{\cA}{{\cal A}}
\nc{\cB}{{\cal B}}
\nc{\cD}{{\cal D}}
\nc{\cE}{{\cal E}}
\nc{\cG}{{\cal G}}
\nc{\cH}{{\cal H}}
\nc{\cL}{{\cal L}}
\nc{\cO}{{\cal O}}
\nc{\cT}{{\cal T}}
\nc{\cN}{{\cal N}}
\nc{\cR}{{\cal R}}
\nc{\rvac}[1]{|{\cal O}#1\rangle}
\nc{\lvac}[1]{\langle{\cal O}#1|}
\nc{\rvacb}[1]{|{\cal O}_\beta #1\rangle}
\nc{\lvacb}[1]{\langle{\cal O}_\beta #1 |}
\nc{\bb}{\bar{\beta}}
\nc{\bt}{\tilde{\beta}}
\nc{\ctH}{\tilde{\cal H}}
\nc{\chH}{\hat{\cal H}}
\nc{\1}{\aa}
\nc{\2}{\"{a}}
\nc{\3}{\"{o}}
\nc{\4}{\AA}
\nc{\5}{\"{A}}
\nc{\6}{\"{O}}
\nc{\al}{\alpha}
\nc{\g}{\gamma}
\nc{\Del}{\Delta}
\nc{\eps}{\epsilon}
\nc{\lam}{\lambda}
\nc{\Om}{\Omega}
\nc{\ve}{\varepsilon}
\nc{\mn}{{\mu\nu}}
\nc{\vp}{\varphi}

\nc{\rf}[1]{(\ref{#1})}
\nc{\nn}{\nonumber \\*}
\nc{\bfB}{\bf{B}}
\nc{\bfv}{\bf{v}}
\nc{\bfx}{\bf{x}}
\nc{\bfy}{\bf{y}}
\nc{\vx}{\vec{x}}
\nc{\vy}{\vec{y}}
\nc{\oB}{\overline{B}}
\nc{\oI}{\overline{I}}
\nc{\oR}{\overline{R}}
\nc{\rar}{\rightarrow}
\nc{\ti}{\times}
\nc{\slsh}{\hskip-5pt/}
\nc{\sm}{Standard~Model~}
\nc{\MP}{M_{\rm Pl}}
\nc{\tp}{t_{\rm Pl}}

\nc{\pmin}{p_{\rm min}}
\nc{\pmax}{p_{\rm max}}
\nc{\fo}{f_0}
\nc{\foi}{f_{0,i}\,}
\nc{\fop}{f_0^P}
\nc{\fou}{f_0^U}

\nc{\eff}{{\rm eff}}
\nc{\MT}{M_{\rm T}}
\nc{\ML}{M_{\rm L}}
\nc{\kk}{\vek{k}}
\nc{\pp}{{\rm p}}
\nc{\half}{{1\over 2}}
\nc{\w}{\omega}
\nc{\uhat}{\hat{U}_\w}

\nc{\etal}{\mbox{\it et al.\,}}
\nc{\ie}{{\it i.e.\,}}
\nc{\eg}{{\it e.g.\,}}
\nc{\trh}{T_{\rm RH}}
\nc{\ad}{{a'\over a}}
\nc{\bd}{{b'\over b}}
\nc{\Rd}{{R'\over R}}
\nc{\diag}{{\textrm{diag}}}
\nc{\mato}[1]{\tilde{#1}}
\nc{\sech}{\textrm{sech}}
\nc{\I}{\textrm{I}}
\nc{\II}{\textrm{II}}
\nc{\III}{\textrm{III}}
\nc{\vev}[1]{\langle #1 \rangle}
\nc{\hyp}{\,\; F_{1{\hskip -16pt}2}{\hskip 11pt}}
\nc{\brhom}{\overline{\rho}_M}
\nc{\brho}{\overline{\rho}}
\nc{\rhob}{\overline{\rho}}
\nc{\Pb}{\overline{P}}
\nc{\bH}{\overline{H}}

\nc{\lcdm}{$\Lambda$CDM }

\def\my{\hbox{\large$\bigcirc$\hspace{-0.58cm}
\raise.5ex\hbox{\tiny{o o}}\kern-.62em
\lower.5ex\hbox{--}}\ }

\nc{\row}{\my\my\my\my\my\my\my}
 
\nc{\mytilde}[1]{{\hskip 2.2pt}$\tilde{}${\hskip -2.2pt}#1}

\begin{document}

\title{Are all modes created equal? An analysis of the WMAP 5- and 7-year data without inflationary prejudice}

\author{Eirik Gjerl{\o}w}
\email{eirik.gjerlow@astro.uio.no}
\affiliation{Institute of theoretical astrophysics, University of Oslo, 
P.O. Box 1029, N-0315 Oslo, NORWAY}
\author{\O ystein Elgar\o y}
\email{oelgaroy@astro.uio.no}
\affiliation{Institute of theoretical astrophysics, University of Oslo, 
P.O. Box 1029, N-0315 Oslo, NORWAY}

\date{\today}

\begin{abstract}
We submit recent claims of a semi-significant detection of primordial tensor perturbations in the WMAP data to a closer 
scrutiny.  Our conclusion is in brief that no such mode is present at a detectable level once the analysis is done more carefully.  These claims have their root in a brief debate in the late 1990s about the standard calculation of the scalar and 
tensor spectra in standard inflationary theory, where Grishchuk and collaborators claimed that their amplitudes should be roughly equal. We give a brief summary of the debate and our own reasons for why the standard calculation is correct.

\end{abstract}

\pacs{98.80.-k, 98.80.Cq, 95.85.Sz}

\preprint{}

\maketitle

\section{Introduction}

Cosmic inflation has become a cornerstone of what may be called the standard model of cosmology. Apart from its 
role in lessening the demands on the need for fine-tuned initial conditions for the observable universe, the major success of the inflationary paradigm has been to explain the existence and statistical properties of the perturbations in the distributions of matter and radiation (see \cite{dodelson} for a gentle introduction).  Inflation, in the simplest versions realized by a self-interacting 
scalar field, predicts a nearly scale-invariant initial power spectrum of density perturbations.  The subsequent evolution of these perturbations limits the 
amount of information we obtain from observations of the statistical distribution of matter in 
the present universe.  This is a pity, since inflation most probably involves physics at energies exceeding those obtainable in particle accelerators by several orders of magnitude, and can 
therefore give important constraints on theories that go beyond the present Standard Model of particle physics.   

Fortunately, another robust prediction of inflation is that it also blows up the quantum 
fluctuations in the spacetime metric to observable scales and produces a spectrum of gravitational waves \cite{dodelson}.  These remain unaffected by the subsequent cosmic evolution, and therefore 
provide a window to the physics of the inflationary process itself.  Unfortunately, the 
standard prediction of single-field inflation is that the amplitudes of these waves are 
much lower than those of the density perturbations, and are hence much harder to observe.  
This standard lore has, however, been challenged, at several times and in various papers, 
by Grishchuk (see e.g. \cite{gris1,gris2}).  His claim is that the standard inflationary treatment of perturbations 
is wrong, and that when the calculation is done correctly what emerges are spectra of 
density (scalar) and gravitational wave (tensor) perturbations with roughly the same amplitude.   Furthermore, in a series of recent papers \cite{zhao1,zhao2,zhao3} he and collaborators claim to 
find weak evidence for a gravitational wave signal in the WMAP 5- and 7-year data \cite{wmap5,wmap7}.  
If his claims are correct, a convincing detection of primordial gravitational waves may 
be very close, perhaps already with Planck data.  They therefore merit closer scrutiny, which 
is what we set out to provide in the present paper.  

The structure of our paper is as follows: In section II we summarize and comment on the 
analyses of the WMAP 5- and 7-year data found in Zhao et al. \cite{zhao2,zhao3}, before giving an outline 
of our strategy in section III.  In section IV we present results from a full analysis 
of the WMAP 7-year data, taking all relevant parameters into account.  This provides 
a background for judging the results of the more restricted analyses of Zhao et al., 
which we carry out improved versions of in section V.  Section VI contains our comments 
on the theoretical aspects of the disagreement on the relation between scalar and 
tensor perturbations.  Finally, we summarize our findings and conclude in section VII.

\section{A summary of the analysis in Zhao et al.}
We will in the current section briefly describe the analyses performed in
\cite{zhao2} and \cite{zhao3}, and the results and conclusions of those analyses, in
order to motivate the analyses performed in the current paper. 

Both papers contain analyses of WMAP data and forecasts for the Planck mission.
Our concern in this paper is the data analysis part, and so we only present the
methodology, results, and conclusions of the WMAP data analyses.

The authors first derive the likelihood function, which depends on the true
$C_{\ell}$'s as well as the observed data points and noise values. They use the
WMAP data points and noise values as provided at the LAMBDA web page 
\footnote{http://lambda.gsfc.nasa.org}, although they approximate the noise values to be uncorrelated between different multipoles.

Using the MCMC method, the authors use the WMAP5
(in \cite{zhao2}) and WMAP7 (in \cite{zhao3}) data to map the likelihood
functions for three parameters: $A_s$ (the amplitude of the scalar perturbations), $n_s$ (the scalar spectral index), and $R$. The first two of these
are the standard cosmological parameters, while $R$ is the tensor-to-scalar
quadrupole ratio, which the authors use instead of the standard parameter
$r$, the tensor-to-scalar power spectrum ratio at $k=k_p = 0.002$ Mpc$^{-1}$. (These are roughly related by $r\approx 2R$ \cite{zhao2}.) 
The other cosmological
parameters are fixed to their mean values from the WMAP team 5-year
\citep{komatsu1} and 7-year \citep{komatsu2} analyses.

The authors only use the multipole ranges $2\leq \ell \leq 100$ and $101 \leq
\ell \leq 220$ in their analyses - the former for the actual analysis, and
the latter in order to show that gravitational waves have little or no effect
beyond $\ell \approx 100$, and that it therefore is futile to use multipoles
higher than this in the search for gravitational waves. Also, they only use the
TT and TE power spectra, saying that the EE and BB modes are not very
informative.

For their $2\leq \ell \leq 100$ analyses, the authors find the following
3-dimensional ML values: 
\begin{widetext}
\begin{equation}
  R = 0.229,\ \ \ \ n_s=1.086,\ \ \ \ A_s=1.920\times10^{-9}
  \label{eq:grish5yrml}
\end{equation}
for the 5-year analysis and
\begin{equation}
  R = 0.264,\ \ \ \ n_s = 1.111, \ \ \ \ A_s = 1.832\times10^{-9}
  \label{eq:grish7yrml}
\end{equation}
for the 7-year analysis. Further, when marginalising the distributions for each
of the parameters, they find the following peak values and Minimum Credible
Interval (MCI) (see \cite{hamann} for an explanation of this interval)
limits:
\begin{equation}
  R = 0.266\pm 0.171,\ \ \ \ n_s = 1.107^{+0.087}_{-0.070},\ \ \ \ A_s =
  (1.768^{+0.307}_{-0.245})\times10^{-9}
  \label{eq:grish5yrpeak}
\end{equation}
for the 5-year analysis and
\begin{equation}
  R = 0.273^{+0.185}_{-0.156},\ \ \ \ n_s = 1.112^{+0.089}_{-0.064},\ \ \ \ A_s =
  (1.765^{+0.279}_{-0.263})\times10^{-9}
  \label{eq:grish7yrpeak}
\end{equation}
for the 7-year analysis, where all limits are the $68.3\%$ confidence intervals.
\end{widetext}

These results contrast with those of the WMAP team, which finds no compelling
evidence for gravitational waves \citep{wmap7}. In order to explain this
discrepancy, Zhao et al. suggest (as we have mentioned) that it is wrong
to use higher multipoles than $\ell=100$ in the search for gravitational waves.
In addition, they point to their $101\leq \ell \leq 220$ analyses, where they
find the following 68.3\% MCI limits for $n_s$: $n_s = 0.948^{+0.052}_{-0.061}$
(5-year-analysis) and $n_s = 0.969^{+0.083}_{-0.063}$ (7-year analysis).
Comparing with the $\ell = 2-100$ analysis, these MCI limits do not overlap (5-year) or
just barely overlaps (7-year) with each other. This
suggests to the authors that the assumption of a $\ell$ (or $k$)-independent
spectral index is erroneous, and this assumption combined with the unwarranted
use of multipoles larger than $100$ is what leads the WMAP team to the wrong
result. They also mention that the WMAP team uses inflationary predictions,
in particular the consistency relation $r = -n_t/8$ (where $n_t$ is the tensor spectral index), in their analyses, which makes the analysis biased from the outset.
\subsection{Preliminary comments}
We here make some short comments about the conclusions reached in the two
papers by Zhao et al.

First of all, we disagree with their claim that when looking for gravitational
waves, one should only include the multipoles that are affected by them. As the
authors themselves have pointed out, there are degeneracies between several
parameters at the low $\ell$ range: $A_s$, $n_s$, and $R$ can be modified so
that two different values of, say, $R$ can produce almost exactly the same power
spectrum at large scales. Hence, it makes sense to include as much data as we
can in the likelihood analysis: Small-scale data may perhaps constrain
$A_s$ or $n_s$, which in turn gives better constraints on $R$, since the
degeneracies may then be broken to some degree. If, when including small-scale
data, we get a small value for $R$, it simply means that there are very few
models where a large $R$ can work together with $n_s$ and $A_s$ (and possibly
other parameters) to produce a
power spectrum that fits both on large and small scales.

Next, concerning the non-compatible values of $n_s$ for $2\leq \ell \leq
100$ and $101\leq \ell \leq 220$: There does indeed seem to be a tension
between these results and the assumption of a constant spectral index. However,
a 1$\sigma$ signal is a hint, not a detection.  Further,
even if the assumption of a constant spectral index turns out to be wrong, it
does not follow that this assumption is what leads to the discrepancy
between their analyses and those of the WMAP team - the solution to the
$n_s$-tension could still yield a negligible $R$, especially considering the
degeneracies mentioned above. More on this below.

Finally, we agree with the authors that the analysis of the WMAP team may be biased
by the explicit use of the consistency relation in their data analysis: This
relation should appear automatically if the theory of single-field inflation is correct, and
not be used as a constraint in the analysis.  However, we also recognise the need to reduce the
number of free parameters in the likelihood analysis, which the WMAP team cites
as their reason for using the consistency relation.

\section{Outline of our strategy}
Our aim in this paper is both to verify the analyses performed in \cite{zhao2} and \cite{zhao3}, but also to further test the conclusions they draw from these analyses. To this end, we will utilise the MCMC method as Zhao et al. do, but our approach will differ somewhat from theirs.

Most importantly, we will be using the CosmoMC software \citep{lewis} to perform the likelihood analysis. The advantages of using this rather than a self-made sampler should be obvious: It is well-tested, and bugs and errors are less likely to occur. The additional advantage is that this software is well integrated with the WMAP likelihood software, so that there is no need for e.g. making approximations to the noise values as Zhao et al. do. All power spectrum modes can be incorporated without extra effort as well. Also, it allows for simulation of the lensing effect on the power spectra, and we use this feature throughout our analyses.

To start off, in section \ref{sec:fullan}, we first do a full analysis of the WMAP7 data for $2\leq\ell\leq\ell_{max}$, where $\ell_{max}$ is the highest multipole for which there are data, and all power spectrum modes. In this analysis we use the constraint $n_s=n_t-1$ instead of the consistency relation, following \cite{zhao1,zhao2,zhao3} where this constraint was imposed for theoretical reasons.  This gives an indication of what the WMAP data would give when the consistency relation is not enforced. 

We analyse the low $\ell$ WMAP data in section \ref{sec:lowl}. First, in section \ref{sec:grishan}, we repeat the analyses performed by Zhao et al. in order to see if the use of the exact noise values and CosmoMC software has any impact on the results. We choose to vary the optical depth to reionization, $\tau$, in addition to the three parameters mentioned above, since this parameter can be expected to share some degeneracy with the three others. We suggest a way to obtain more appropriate best-fit values at which to fix the parameters we do not vary, and repeat the analyses using these best-fit values.

Taking our low-$\ell$ analysis a step further, in section \ref{sec:ssan} we aim to test the claim made by Zhao et al. that the assumption of a constant $n_s$ is what leads the WMAP team to overlook the gravitational wave contribution to the CMB. To do this, we introduce a 'step-like' spectral index with a jump at $\ell =100$, and do analyses for the range $\ell=2-220$.
\section{Full analysis of all WMAP7 data, with all relevant parameters}
\label{sec:fullan}
In this section, we do an initial full analysis of the WMAP 7-year data. We use, as already mentioned, the CosmoMC sampler to find the likelihood function for a 7-dimensional parameter space, consisting of the parameters $\Omega_{\rm b}h^2$, $\Omega_{\rm DM}h^2$, $\tau$, $n_{\rm s}$, $\log(10^{10}A_{\rm s})$, $\theta$, and $R$. All these parameters are standard CosmoMC parameters ($\Omega_{\rm b}h^2$ is the physical baryon density, $\Omega_{\rm DM}h^2$ is the physical density of cold dark matter), except $R$, with $\theta$ being the ratio of the sound horizon to the angular diameter distance, which is being used instead of the Hubble parameter because it is less correlated with other parameters. We modify CosmoMC to use $R$ instead of $r$. As also mentioned before, we use the relation $n_t=n_s-1$ instead of the inflationary consistency relation. Following Zhao et al. and the WMAP team, we use $k_p = 0.002$ Mpc$^{-1}$ as the pivot wavenumber.

This choice of parameters implies certain assumptions - namely, a flat Universe, dark energy with the equation of state $w=-1$, massless neutrinos, and no running of the spectral index.

We do not use a fixed number of samples; since we do 8 chains simultaneously, we use the CosmoMC feature to check for convergence of confidence limits: By computing the variance between chains of 2.5 per cent of the distribution tails, the error must be smaller than 0.2 in units of the distribution's standard deviation. This means that we can be sure that the sample confidence limits are not too far away from the true confidence limits.

We use the four power spectrum modes $TT$, $TE$, $EE$, and $BB$, and we do the analysis for the range $2\leq \ell \leq \ell_{max}$, where $\ell_{max}$ is 1200 for the $TT$ mode, 800 for the $TE$ mode, and 23 for the $EE$ and $BB$ modes.
\subsection{Results}
The results of the analysis described above are shown in table \ref{tab:fullan}. Both full-dimensional ML values and one-dimensional peaks and MCI limits are shown.
\begin{table}
    \centering
  \renewcommand\arraystretch{1.2}
    \begin{tabular}{ccc}
        Parameter & ML value & One-dimensional peaks and MCI limits \\
        & & $X\ ^{68\%\uparrow,\ 95\%\uparrow}_{68\%\downarrow,\
        95\%\downarrow}$ \\
        \hline
        $\Omega_bh^2$ & 0.0227 & 0.0232 $_{0.0224,\ 0.0217}^{0.0239,\ 0.0248}$ \\
        $\Omega_{DM}h^2$ & 0.1107 & 0.1065 $_{0.1003,\ 0.0936}^{0.1130,\ 0.1189}$\\
        $\theta$ & 1.040 & 1.041 $_{1.038,\ 1.035}^{1.044,\ 1.047}$ \\
        $\tau$ & 0.0880 & 0.0894 $_{0.0738,\ 0.0611}^{0.1051,\ 0.1229}$\\
        $n_s$ & 0.970 & 0.982 $_{0.964,\ 0.947}^{1.007,\ 1.034}$ \\
        $\log(10^{10}A_s)$ & 3.174 & 3.130 $_{3.052,\ 2.958}^{3.195,\ 3.249}$ \\
        $R$ & 0.012 & 0.010 $_{0.000,\ 0.000}^{0.089,\ 0.174}$\\
        \hline
    \end{tabular}
    \caption{ML points and marginalised one-dimensional peaks and MCI limits
    for a likelihood analysis for seven parameters using the WMAP 7-year data. MCI limits are
    given with both 68\% and 95\% confidence.}
    \label{tab:fullan}
\end{table}
\subsection{Discussion}
We have performed a full-scale likelihood analysis of the WMAP 7-year data, not using the consistency relation, in accord with the complaints voiced by Zhao et al. Despite this, there is little reason to claim any detection of gravitational waves in the CMB, as the $R=0$ hypothesis is well within confidence bounds. This means that the only explanation left, as offered by Zhao et al., to explain the discrepancy between their results and those of the WMAP team must be the (erronously?) assumed constancy of the spectral index for all $\ell$'s. We will test this explanation presently, but first we will repeat their analyses of the low $\ell$ WMAP data.
\section{Low-{$\ell$} analysis}
\label{sec:lowl}
We will now focus on the large-scale WMAP data, as the central claims of Zhao et al. pertains to these scales. First, we repeat the work of Zhao et al. in section \ref{sec:grishan} before we move on to testing the authors' claim concerning the non-constancy of the spectral index in section \ref{sec:ssan}.

We will need to limit the number of free parameters to use in this section, as we will only be using multipoles up to $\ell=220$, and this data range contains too little information to be able to constrain all parameters at once. Which parameters we vary will be specified below. We will, however, never use the Hubble parameter, instead using $\theta$. This then differs from the approach of Zhao et al.

The parameters we do not vary must be given a certain value. As mentioned above, Zhao et al. have chosen to use the mean value of the marginalised one-dimensional distributions for each parameter from the WMAP standard analysis. We find two problems with this approach: First, the WMAP standard analysis excluded gravitational waves, making the best-fit parameters obtained less valid when we want to include gravitational waves in our analyses. Second, we think that the mean values are not the best values to use. The one-dimensional peak values would be better, as these represent the most probable value of each parameter given the data we have (of course, for Gaussian distributions, the mean and peak values coincide). In our opinion it is better to choose the parameter values corresponding to the peak of the full, multi-dimensional likelihood function: These taken together are the closest to what one may call the Standard Model.

In order to remedy these issues, we here do the following: We do an initial analysis of most available data: WMAP 7-year data ($2\leq \ell\leq\ell_{max}$), matter power spectrum data (SDSS, fourth data release \citep{adelman}), supernova observations (SDSS, ESSENCE \citep{miknaitis}, SNLS \citep{balland}, HST \citep{riess} and various low-redshift supernova observations), Lyman-alpha data (SDSS), and priors on the age of the Universe, the Hubble parameter (from the HST), and $\Omega_b$ (from BBN). We let vary the same parameters as in section \ref{sec:fullan}, but use $r$ instead of $R$. We do this as $R$ is harder to implement for non-CMB data, since it is just the ratio of the tensor contribution to the CMB quadrupole to the scalar contribution, while $r$ is more 'fundamental', as it describes the tensor/scalar ratio of the primordial power spectra. We do keep using the relation $n_t=n_s-1$. We then take the full-dimensional ML values from this analysis and use these as best-fit values for the rest of the analyses (though we will also use the best-fit values used by Zhao et al., see below).

The results of this analysis are shown in table \ref{tab:compan}.
\begin{table}
    \centering
  \renewcommand\arraystretch{1.2}
    \begin{tabular}{ccc}
        Parameter & ML value & One-dimensional peaks and MCI limits \\
        & & $X\ ^{68\%\uparrow,\ 95\%\uparrow}_{68\%\downarrow,\
        95\%\downarrow}$ \\
        \hline
        $\Omega_bh^2$ & 0.0226 & 0.0228 $_{0.0223,\ 0.0218}^{0.0233,\ 0.0238}$ \\
        $\Omega_{DM}h^2$ & 0.1177 & 0.1176 $_{0.1145,\ 0.1118}^{0.1203,\ 0.1229}$\\
        $\theta$ & 1.040 & 1.040 $_{1.038,\ 1.036}^{1.043,\ 1.045}$ \\
        $\tau$ & 0.0810 & 0.0876 $_{0.0749,\ 0.0625}^{0.1028,\ 0.1176}$\\
        $n_s$ & 0.963 & 0.969 $_{0.957,\ 0.945}^{0.983,\ 0.996}$ \\
        $\log(10^{10}A_s)$ & 3.207 & 3.203 $_{3.165,\ 3.123}^{3.243,\ 3.281}$ \\
        $r$ & 0.015 & 0.000 $_{0.000,\ 0.000}^{0.076,\ 0.168}$\\
        \hline
    \end{tabular}
    \caption{ML points and marginalised one-dimensional peaks and MCI limits
    for a likelihood analysis for seven parameters, taking into account CMB,
    power spectrum, and supernova data, in addition to various prior
    constraints on some parameters, described in the text. MCI limits are
    given with both 68\% and 95\% confidence.}
    \label{tab:compan}
\end{table}
We will from here on carry out two versions of the analysis: One using the ML values found in table \ref{tab:compan}, and one using the best-fit values used by Zhao et al., in order to examine whether using different best-fit values makes a large difference in the results. When using the best-fit values from Zhao et al., we use the 5-year values for the 5-year analysis, and 7-year values for the 7-year analysis. We will, as mentioned, not use the Hubble parameter, so in this we do not follow Zhao et al. The value for $\theta$ will always be the ML value from table \ref{tab:compan}.
\subsection{Analysis with CosmoMC}
\label{sec:grishan}
In this section, we repeat the low-$\ell$ analyses in \cite{zhao2} and \cite{zhao3} with certain elaborations and modifications, which we here specify:

As before, we use CosmoMC with the WMAP likelihood software, and include all power spectrum modes in the analysis. We do analyses for both 5-year and 7-year WMAP data in order to compare with both papers by Zhao et al. We include $\tau$ as a parameter to be varied, though we do analyses where we vary only the three original parameters as well. We do all analyses over three ranges: $2\leq \ell \leq100$, $101\leq\ell\leq220$, and $2\leq\ell\leq220$. (The $\ell=101-220$ contained too little information to constrain four parameters at once, so we did not include $\tau$ in the analyses for this range.) We do an additional run where we fix $R$ to be its full-dimensional ML value from the $\ell =2-100$ analysis where $\tau$ is not varied. This run is done for later comparison.

\subsubsection{Results}
The results of the above analyses are shown in tables \ref{tab:grishannml} (for the revised best-fit values) and \ref{tab:grishangml} (for the best-fit values used by Zhao et al.) We have assigned to each run a number for easier referral.
\begin{table*}
    \centering
  \renewcommand\arraystretch{1.2}
  \begin{tabular}{ccccc}
    Parameter & \multicolumn{2}{c}{ML value} &
    \multicolumn{2}{c}{One-dimensional peaks and MCI limits} \\
& & & \multicolumn{2}{c}{$X\ ^{68\%\uparrow,\ 95\%\uparrow}_{68\%\downarrow,\
95\%\downarrow}$} \\
      & 5-year & 7-year & 5-year & 7-year \\
      \hline
      \multicolumn{5}{c}{Run 1 ($\ell=2-100$, varying $n_s$, $A_s$, and $R$)} \\
      \hline
      $n_s$ & 1.038 & 1.036 & 1.047 $^{1.126,\ 1.201}_{0.994,\ 0.949}$ &
      1.049 $^{1.112,\ 1.186}_{0.996,\ 0.949}$ \\
      $\log(10^{10} A_s)$ & 3.066 & 3.067 & 3.047 $^{3.149,\ 3.220}_{2.896,\
      2.738}$ &
      3.050 $^{3.145,\ 3.217}_{2.916,\ 2.775}$ \\
      $R$ & 0.104 & 0.107 & 0.069 $^{0.251,\ 0.466}_{0.000,\ 0.000}$ &
      0.138 $^{0.229,\ 0.423}_{0.000,\ 0.000}$ \\
      \hline
      \multicolumn{5}{c}{Run 2 ($\ell=101-220$, varying $n_s$, $A_s$, and
      $R$)} \\
      \hline
      $n_s$ & 0.925 & 0.953 & 0.961 $^{1.031,\ 1.104}_{0.893,\ 0.828}$ &
      0.979 $^{1.062,\ 1.138}_{0.914,\ 0.846}$ \\
      $\log(10^{10} A_s)$ & 3.291 & 3.234 & 3.203 $^{3.355,\ 3.489}_{3.066,\
      2.909}$ &
      3.164 $^{3.316,\ 3.455}_{3.005,\ 2.844}$ \\
      $R$ & 0.028 & 0.245 & 0.335 $^{1.044,\ 2.050}_{0.000,\ 0.000}$ &
      0.112 $^{1.144,\ 2.161}_{0.000,\ 0.000}$ \\
      \hline
      \multicolumn{5}{c}{Run 3 ($\ell=2-220$, varying $n_s$, $A_s$, and $R$)} \\
      \hline
      $n_s$ & 0.976 & 0.991 & 0.998 $^{1.034,\ 1.073}_{0.973,\ 0.949}$ &
      1.010 $^{1.048,\ 1.091}_{0.980,\ 0.955}$ \\
      $\log(10^{10} A_s)$ & 3.185 & 3.158 & 3.136 $^{3.191,\ 3.237}_{3.068,\
      2.984}$ &
      3.116 $^{3.179,\ 3.229}_{3.042,\ 2.951}$ \\
      $R$ & 0.001 & 0.021 & 0.000 $^{0.100,\ 0.210}_{0.000,\ 0.000}$ &
      0.000 $^{0.121,\ 0.243}_{0.000,\ 0.000}$ \\
      \hline
      \multicolumn{5}{c}{Run 4 ($\ell=2-220$, varying $n_s$ and
      $A_s$)} \\
      \hline
      $n_s$ & 1.018 & 1.024 & 1.014 $_{0.999,\ 0.980}^{1.039,\ 1.056}$ &
      1.021 $_{1.004,\ 0.984}^{1.044,\ 1.065}$ \\
      $\log(10^{10} A_s)$ & 3.098 & 3.089 & 3.098 $_{3.057,\ 3.026}^{3.134,\
      3.172}$ & 3.092 $_{3.050,\ 3.009}^{3.127,\ 3.164}$ \\
      \hline
      \multicolumn{5}{c}{Run 5 ($\ell=2-100$, varying $\tau$, $n_s$, $A_s$, and
      $R$)} \\
      \hline
      $\tau$ & 0.0975 & 0.0941 & 0.1021 $^{0.1217,\ 0.1454}_{0.0798,\ 0.0624}$ &
      0.0937 $^{0.1128,\ 0.1312}_{0.0791,\ 0.0638}$ \\
      $n_s$ & 1.076 & 1.063 & 1.095 $^{1.174,\ 1.247}_{1.028,\ 0.969}$ &
      1.084 $^{1.145,\ 1.221}_{1.016,\ 0.967}$ \\
      $\log(10^{10} A_s)$ & 3.025 & 3.041 & 2.982 $^{3.122,\ 3.209}_{2.862,\
      2.717}$ &
      3.028 $^{3.123,\ 3.212}_{2.892,\ 2.760}$ \\
      $R$ & 0.167 & 0.146 & 0.208 $^{0.341,\ 0.520}_{0.044,\ 0.000}$ &
      0.136 $^{0.293,\ 0.447}_{0.038,\ 0.000}$ \\
      \hline
      \multicolumn{5}{c}{Run 6 ($\ell=2-220$, varying $\tau$, $n_s$, $A_s$, and
      $R$)} \\
      \hline
      $\tau$ & 0.0876 & 0.0876 & 0.0869 $_{0.0723,\ 0.0573}^{0.1079,\ 0.1276}$ &
      0.0879 $_{0.0747,\ 0.0616}^{0.1056,\ 0.1227}$ \\
      $n_s$ & 0.980 & 0.993 & 1.008 $_{0.977,\ 0.952}^{1.044,\ 1.088}$ &
      1.020 $_{0.985,\ 0.959}^{1.056,\ 1.102}$ \\ 
      $\log(10^{10} A_s)$ & 3.191 & 3.166 & 3.139 $_{3.068,\ 2.982}^{3.199,\
      3.245}$ & 3.134 $_{3.047,\ 2.955}^{3.186,\ 3.235}$ \\
      $R$ & 0.002 & 0.022 & 0.000 $_{0.000,\ 0.000}^{0.109,\ 0.225}$ &
      0.015 $_{0.000,\ 0.000}^{0.128,\ 0.249}$ \\
      \hline
      \multicolumn{5}{c}{Run 7 ($\ell=2-220$, varying $\tau$, $n_s$, and
      $A_s$)} \\
      \hline
      $\tau$ & 0.0882 & 0.0912 & 0.0942 $_{0.0709,\ 0.0568}^{0.1079,\ 0.1290}$ &
      0.0927 $_{0.0759,\ 0.0622}^{0.1058,\ 0.1206}$ \\
      $n_s$ & 1.020 & 1.028 & 1.021 $_{1.003,\ 0.981}^{1.045,\ 1.066}$ &
      1.030 $_{1.008,\ 0.988}^{1.051,\ 1.071}$ \\
      $\log(10^{10} A_s)$ & 3.108 & 3.101 & 3.101 $_{3.065,\ 3.026}^{3.149,\
      3.191}$ & 3.094 $_{3.057,\ 3.017}^{3.137,\ 3.176}$ \\
      \hline
  \end{tabular}
  \caption[Results from repeated analysis]{The
  unmarginalised ML points and
  one-dimensional peaks and MCI limits resulting from a likelihood analysis using both
  5-year and 7-year $WMAP$ data for various $\ell$ ranges, and with various
  parameters. The best-fit values for the fixed parameters are the full-dimensional ML values taken from table \ref{tab:fullan}.
  MCI limits are given
  with both 68\% and 95\% confidence.}
  \label{tab:grishannml}
\end{table*}

\begin{table*}
    \centering
  \renewcommand\arraystretch{1.2}
  \begin{tabular}{ccccc}
    Parameter & \multicolumn{2}{c}{ML value} &
    \multicolumn{2}{c}{One-dimensional peaks and MCI limits} \\
& & & \multicolumn{2}{c}{$X\ ^{68\%\uparrow,\ 95\%\uparrow}_{68\%\downarrow,\
95\%\downarrow}$} \\
      & 5-year & 7-year & 5-year & 7-year \\
      \hline
      \multicolumn{5}{c}{Run 1 ($\ell=2-100$, varying $n_s$, $A_s$, and $R$)} \\
      \hline
      $n_s$ & 1.042 & 1.046 & 1.063 $^{1.133,\ 1.212}_{0.997,\ 0.951}$ &
      1.062 $^{1.125,\ 1.194}_{0.999,\ 0.949}$ \\
      $\log(10^{10} A_s)$ & 3.042 & 3.036 & 2.997 $^{3.130,\ 3.204}_{2.870,\
      2.706}$ &
      2.974 $^{3.129,\ 3.206}_{2.889,\ 2.743}$ \\
      $R$ & 0.120 & 0.129 & 0.140 $^{0.259,\ 0.472}_{0.000,\ 0.000}$ &
      0.132 $^{0.261,\ 0.427}_{0.022,\ 0.000}$ \\
      \hline
      \multicolumn{5}{c}{Run 2 ($\ell=101-220$, varying $n_s$, $A_s$, and
      $R$)} \\
      \hline
      $n_s$ & 0.921 & 0.960 & 0.965 $^{1.027,\ 1.097}_{0.893,\ 0.834}$ &
      0.994 $^{1.063,\ 1.138}_{0.921,\ 0.859}$ \\
      $\log(10^{10} A_s)$ & 3.282 & 3.207 & 3.210 $^{3.337,\ 3.457}_{3.057,\
      2.908}$ &
      3.128 $^{3.285,\ 3.414}_{2.988,\ 2.830}$ \\
      $R$ & 0.046 & 0.365 & 0.000 $^{0.983,\ 1.978}_{0.000,\ 0.000}$ &
      0.582 $^{1.128,\ 2.083}_{0.000,\ 0.000}$ \\
      \hline
      \multicolumn{5}{c}{Run 3 ($\ell=2-220$, varying $n_s$, $A_s$, and $R$)} \\
      \hline
      $n_s$ & 0.976 & 0.997 & 0.998 $^{1.036,\ 1.080}_{0.969,\ 0.945}$ &
      1.012 $^{1.048,\ 1.093}_{0.978,\ 0.953}$ \\
      $\log(10^{10} A_s)$ & 3.169 & 3.132 & 3.124 $^{3.183,\ 3.229}_{3.047,\
      2.956}$ &
      3.098 $^{3.171,\ 3.220}_{3.030,\ 2.937}$ \\
      $R$ & 0.009 & 0.043 & 0.000 $^{0.113,\ 0.229}_{0.000,\ 0.000}$ &
      0.038 $^{0.126,\ 0.248}_{0.000,\ 0.000}$ \\
      \hline
      \multicolumn{5}{c}{Run 4 ($\ell=2-220$, varying $n_s$ and
      $A_s$)} \\
      \hline
      $n_s$ & 1.021 & 1.031 & 1.022 $_{1.001,\ 0.982}^{1.040,\ 1.059}$ &
      1.031 $_{1.011,\ 0.991}^{1.053,\ 1.072}$ \\
      $\log(10^{10} A_s)$ & 3.075 & 3.062 & 3.073 $_{3.041,\ 3.003}^{3.115,\
      3.151}$ & 3.057 $_{3.018,\ 2.983}^{3.100,\ 3.139}$ \\
      \hline
      \multicolumn{5}{c}{Run 5 ($\ell=2-100$, varying $\tau$, $n_s$, $A_s$, and
      $R$)} \\
      \hline
      $\tau$ & 0.0965 & 0.0966 & 0.1002 $^{0.1219,\ 0.1463}_{0.0795,\ 0.0629}$ &
      0.0960 $^{0.1140,\ 0.1334}_{0.0800,\ 0.0661}$ \\
      $n_s$ & 1.069 & 1.060 & 1.081 $^{1.171,\ 1.250}_{1.021,\ 0.965}$ &
      1.078 $^{1.147,\ 1.220}_{1.010,\ 0.959}$ \\
      $\log(10^{10} A_s)$ & 3.017 & 3.027 & 2.977 $^{3.108,\ 3.197}_{2.844,\
      2.694}$ &
      3.006 $^{3.122,\ 3.201}_{2.878,\ 2.739}$ \\
      $R$ & 0.158 & 0.152 & 0.201 $^{0.344,\ 0.509}_{0.059,\ 0.000}$ &
      0.116 $^{0.285,\ 0.445}_{0.032,\ 0.000}$ \\
      \hline
      \multicolumn{5}{c}{Run 6 ($\ell=2-220$, varying $\tau$, $n_s$, $A_s$, and
      $R$)} \\
      \hline
      $\tau$ & 0.0869 & 0.0886 & 0.0924 $_{0.0739,\ 0.0573}^{0.1102,\ 0.1295}$ &
      0.0912 $_{0.0768,\ 0.0630}^{0.1079,\ 0.1232}$ \\
      $n_s$ & 0.976 & 0.997 & 1.005 $_{0.972,\ 0.945}^{1.044,\ 1.090}$ &
      1.011 $_{0.985,\ 0.956}^{1.056,\ 1.099}$ \\ 
      $\log(10^{10} A_s)$ & 3.175 & 3.137 & 3.128 $_{3.050,\ 2.959}^{3.189,\
      3.241}$ & 3.117 $_{3.026,\ 2.947}^{3.164,\ 3.219}$ \\
      $R$ & 0.004 & 0.041 & 0.000 $_{0.000,\ 0.000}^{0.116,\ 0.236}$ &
      0.040 $_{0.000,\ 0.000}^{0.129,\ 0.243}$ \\
      \hline
      \multicolumn{5}{c}{Run 7 ($\ell=2-220$, varying $\tau$, $n_s$, and
      $A_s$)} \\
      \hline
      $\tau$ & 0.0912 & 0.0932 & 0.0905 $_{0.0739,\ 0.0594}^{0.1090,\ 0.1284}$ &
      0.0917 $_{0.0771,\ 0.0636}^{0.1075,\ 0.1247}$ \\
      $n_s$ & 1.023 & 1.035 & 1.027 $_{1.003,\ 0.983}^{1.046,\ 1.067}$ &
      1.037 $_{1.013,\ 0.992}^{1.057,\ 1.078}$ \\
      $\log(10^{10} A_s)$ & 3.085 & 3.067 & 3.086 $_{3.041,\ 3.001}^{3.128,\
      3.168}$ & 3.072 $_{3.024,\ 2.986}^{3.107,\ 3.149}$ \\
      \hline
  \end{tabular}
  \caption[Results from repeated analysis]{The
  unmarginalised ML points and
  one-dimensional peaks and MCI limits resulting from a likelihood analysis using both
  5-year and 7-year $WMAP$ data for various $\ell$ ranges, and with various
  parameters. The best-fit values for the fixed parameters are the same as those used by Zhao et al.  MCI limits are given
  with both 68\% and 95\% confidence.}
  \label{tab:grishangml}
\end{table*}
\subsubsection{Discussion}
From tables \ref{tab:grishannml} and \ref{tab:grishangml}, we note the following features of our analyses: The values for $R$ are consistently lower than those found in the analyses of Zhao et al., and the one-dimensional distributions of $n_s$ for runs 1 and 2 overlap within a 1$\sigma$ interval. This seems to hold no matter which best-fit values we use (though with the best-fit values used by Zhao et al. we find a slightly higher $R$ than with the best-fit values derived here), which indicates that the difference in results from the analysis of Zhao et al. most likely is due to use of the WMAP likelihood software with exact noise values (The use of the $EE$ and $BB$ modes should have less of an impact due to the high noise in these modes).

The inclusion of $\tau$ actually seems to raise the values for $R$ slightly, and we see what Zhao et al. also reported: $R$ has higher values in the 7-year analyses than in the 5-year ones (mostly). Also, in some cases, we do find lower limits for $R$. However, these are very small, and we only have a 1$\sigma$ lower limit, so it is hard to draw any conclusions from these results.

The fact that the distributions for $n_s$ overlap within 1$\sigma$ between runs 1 and 2 weakens the claim that the spectral index really is dependent on $\ell$. Even though the overlap is slight, we must again be reminded that a 1$\sigma$ detection is very weak, statistically speaking. Nevertheless, in the next section we will try to put the claims about $n_s$ made by Zhao et al. to the test.
\subsection{Testing the hypothesis of an $\ell$-dependent $n_s$}
\label{sec:ssan}
In this section, we wish to test, on a very basic level, whether introducing an $\ell$-dependent spectral index gives better fit to data, and whether it may yield a high value for $R$ over a larger multipole range than $\ell=2-100$.

In order to accomplish this, we implement a 'step-like' spectral index: The spectral index is allowed two values - $n_s^{\ell<100}$ below $\ell=100$ and $n_s^{\ell>100}$ above this multipole. This is not implemented directly - rather, $\ell=100$ is translated into $\vec{k}$-space through the relation \citep{elgaroy} $\ell\approx kd_a$, where
\begin{equation}
  d_a \approx \frac{2c}{H_0\Omega_{\rm m}^{0.4}}
  \label{eq:angdist}
\end{equation}
is the angular diameter distance, and $\Omega_{\rm m}=\Omega_{\rm DM}+\Omega_{\rm b}$ is the density parameter of dust. We choose to work in $\vec{k}$-space both because it is easier to implement, and because it makes slightly more sense physically. However, we emphasise that there is no physical justification for such a spectral index - we are just attempting to test a claim which will eventually need some physical justification if found to be probable. 

We still use the relation $n_t=n_s-1$, so that both the scalar and tensor spectral indices attain this step-like behaviour.

Another, more straightforward way to implement an $\ell$-dependence would be to use a simple running of the spectral index. However, we have chosen not to do this because this case is discussed in Zhao et al. and claimed to be unwarranted. Also, the method outlined above is a more direct test of the claims of Zhao et al.

What do we expect from analyses involving such a spectral index, if the claims of Zhao et al. are true? First of all, we would expect that $n_s^{\ell<100}$ and $n_s^{\ell>100}$ converge to the values found in runs 1 (or 5) and 2 in the previous analysis, respectively, if we do a run over the range $\ell=2-220$. Further, we would expect that the values of $R$ found over such a multipole range would be greater than those values found in runs 3 and 6 in the previous analysis. Finally, since we are effectively introducing a new parameter, we would expect a significantly better fit to data - that is, the likelihood values for the ML model of this analysis should be well above the likelihood values of the ML models of runs 3 and 6 in the previous analysis.

We will vary the same parameters as in the previous analysis, both with and without $\tau$. In addition, we will do runs where we fix $R$ (to see if this makes $n_s^{\ell<100}$ and $n_s^{\ell>100}$ converge better to their expected values), and runs where we fix $n_s^{\ell<100}$ and $n_s^{\ell>100}$ (to see if $R$ converges better to its expected value). When fixing $R$ and $n_s^{\ell<100}$, we fix them to their full-dimensional ML value found in run 1, and when fixing $n_s^{\ell>100}$, we fix it to its full-dimensional ML value found in run 2.

We will also do a full analysis, similar to the one in section \ref{sec:fullan}, the only difference being the use of the step-like spectral index. We only do this analysis for the 7-year data, since the analysis in section \ref{sec:fullan} was only carried out for this data set.

The data used are again the WMAP 5-year and 7-year $TT, TE, EE$, and $BB$ data, and again we do all runs twice, once for each set of best-fit parameters.
\subsubsection{Results}
The results of our analyses are shown in tables \ref{tab:ssannml} (our best-fit values) and \ref{tab:ssangml} (the best-fit values of Zhao et al.). The results of the full analysis are shown in table \ref{tab:ssfullan}. Finally, we show the comparison of likelihood values for the ML models in tables \ref{tab:likesnml} (our best-fit values) and \ref{tab:likesgml} (best-fit values used by Zhao et al.).
\begin{table*}
    \centering
  \renewcommand\arraystretch{1.2}
  \begin{tabular}{ccccc}

    Parameter & \multicolumn{2}{c}{ML value} &
    \multicolumn{2}{c}{One-dimensional peaks and MCI limits} \\
& & & \multicolumn{2}{c}{$X\ ^{68\%\uparrow,\ 95\%\uparrow}_{68\%\downarrow,\
95\%\downarrow}$} \\
      & 5-year & 7-year & 5-year & 7-year \\
      \hline
      \multicolumn{5}{c}{Run 8 (varying $n_s^{\ell<100}, n_s^{\ell>100}$, $A_s$, and
      $R$)} \\
      \hline
      $n_s^{\ell<100}$ & 0.993 & 1.007 & 1.018 $_{0.943,\ 0.881}^{1.077,\ 1.142}$ &
      1.021 $_{0.952,\ 0.887}^{1.084,\ 1.148}$ \\
      $n_s^{\ell>100}$ & 0.983 & 0.987 & 1.004 $_{0.973,\ 0.949}^{1.037,\ 1.076}$ &
      1.016 $_{0.982,\ 0.955}^{1.049,\ 1.092}$ \\
      $\log(10^{10} A_s)$ & 3.172 & 3.165 & 3.124 $_{3.060,\ 2.978}^{3.189,\
      3.234}$ &
      3.117 $_{3.039,\ 2.947}^{3.176,\ 3.229}$ \\
      $R$ & 0.004 & 0.004 & 0.000 $_{0.000,\ 0.000}^{0.105,\ 0.231}$ & 0.015
      $_{0.000,\ 0.000}^{0.125,\ 0.262}$ \\
      \hline
      \multicolumn{5}{c}{Run 9 (varying $\tau$, $n_s^{\ell<100}$, $n_s^{\ell>100}$,
      $A_s$, and $R$)} \\
      \hline
      $\tau$ & 0.0824 & 0.0926 & 0.0874 $_{0.0730,\ 0.0561}^{0.1090,\ 0.1286}$
      & 0.0900 $_{0.0762,\ 0.0627}^{0.1076,\ 0.1237}$ \\
      $n_s^{\ell<100}$ & 1.004 & 1.027 & 1.021 $_{0.956,\ 0.888}^{1.093,\ 1.160}$ &
      1.044 $_{0.962,\ 0.896}^{1.111,\ 1.181}$ \\
      $n_s^{\ell>100}$ & 0.981 & 0.997 & 1.004 $_{0.978,\ 0.950}^{1.048,\ 1.090}$ &
      1.020 $_{0.989,\ 0.957}^{1.063,\ 1.109}$ \\
      $\log(10^{10} A_s)$ & 3.176 & 3.169 & 3.138 $_{3.065,\ 2.979}^{3.198,\
      3.250}$ & 3.124 $_{3.038,\ 2.944}^{3.182,\ 3.236}$ \\
      $R$ & 0.003 & 0.017 & 0.000 $_{0.000,\ 0.000}^{0.107,\ 0.237}$ & 0.000
      $_{0.000,\ 0.000}^{0.131,\ 0.274}$ \\
      \hline
      \multicolumn{5}{c}{Run 10 (varying $\tau$, $n_s^{\ell<100}$, $n_s^{\ell>100}$, and
      $A_s$)} \\
      \hline
      $\tau$ & 0.0894 & 0.0924 & 0.0906 $_{0.0725,\ 0.0573}^{0.1086,\ 0.1283}$
      & 0.0929 $_{0.0761,\ 0.0624}^{0.1069,\ 0.1240}$ \\
      $n_s^{\ell<100}$ & 1.025 & 1.039 & 1.031 $_{0.960,\ 0.893}^{1.096,\ 1.162}$ &
      1.025 $_{0.966,\ 0.898}^{1.105,\ 1.178}$ \\
      $n_s^{\ell>100}$ & 1.022 & 1.032 & 1.021 $_{0.999,\ 0.975}^{1.049,\ 1.073}$ &
      1.034 $_{1.005,\ 0.980}^{1.055,\ 1.080}$ \\
      $\log(10^{10}A_s)$ & 3.105 & 3.093 & 3.109 $_{3.061,\ 3.014}^{3.155,\
      3.197}$ & 3.106 $_{3.052,\ 3.006}^{3.146,\ 3.189}$ \\
      \hline
      \multicolumn{5}{c}{Run 11 (varying $\tau$, $A_s$, and
      $R$)} \\
      \hline
      $\tau$ & 0.0726 & 0.0830 & 0.0722 $_{0.0588,\ 0.0460}^{0.0868,\ 0.1010}$
      & 0.0825 $_{0.0687,\ 0.0554}^{0.0943,\ 0.1067}$ \\
      $\log(10^{10} A_s)$ & 3.257 & 3.230 & 3.260 $_{3.228,\ 3.200}^{3.285,\
      3.312}$ &
      3.226 $_{3.199,\ 3.171}^{3.251,\ 3.275}$ \\
      $R$ & $4\times 10^{-5}$ & $2\times 10^{-4}$ & 0.000 $_{0.000,\
      0.000}^{0.011,\ 0.029}$ & 0.000 $_{0.000,\ 0.000}^{0.018,\ 0.043}$ \\
      \hline
      \multicolumn{5}{c}{Run 12 (varying $n_s^{\ell<100}$, $n_s^{\ell>100}$, and $A_s$)} \\
      \hline
      $n_s^{\ell<100}$ & 1.009 & 1.022 & 1.013 $_{0.943,\ 0.882}^{1.076,\ 1.141}$ &
      1.019 $_{0.952,\ 0.894}^{1.080,\ 1.153}$ \\
      $n_s^{\ell>100}$ & 1.017 & 1.024 & 1.018 $_{0.994,\ 0.973}^{1.038,\ 1.060}$ &
      1.022 $_{1.000,\ 0.979}^{1.046,\ 1.068}$ \\
      $\log(10^{10} A_s)$ & 3.101 & 3.089 & 3.100 $_{3.058,\ 3.012}^{3.148,\
      3.189}$ & 3.088 $_{3.045,\ 3.000}^{3.137,\ 3.179}$ \\
      \hline
      \multicolumn{5}{c}{Run 13 (varying $A_s$ and $R$)} \\
      \hline
      $\log(10^{10} A_s)$ & 3.271 & 3.225 & 3.274 $_{3.262,\ 3.254}^{3.280,\
      3.288}$ &
      3.223 $_{3.215,\ 3.207}^{3.232,\ 3.240}$ \\
      $R$ & $8\times10^{-6}$ & $2\times10^{-5}$ & 0.000 $_{0.000,\ 0.000}^{0.012,\ 0.031}$ &
      0.000 $_{0.000,\ 0.000}^{0.019,\ 0.044}$ \\
      \hline
  \end{tabular}
  \caption{Unmarginalised ML points and
  one-dimensional peaks and MCI limits resulting from a likelihood analysis with an $\ell$-dependent $n_s$, using the best-fit values derived in this paper. Results from the 5-year and 7-year analyses are shown, and all runs were over the range
  $\ell=2-220$.
  MCI limits are given
  with both 68\% and 95\% confidence.}
  \label{tab:ssannml}
\end{table*}
\begin{table*}
    \centering
  \renewcommand\arraystretch{1.2}
  \begin{tabular}{ccccc}

    Parameter & \multicolumn{2}{c}{ML value} &
    \multicolumn{2}{c}{One-dimensional peaks and MCI limits} \\
& & & \multicolumn{2}{c}{$X\ ^{68\%\uparrow,\ 95\%\uparrow}_{68\%\downarrow,\
95\%\downarrow}$} \\
      & 5-year & 7-year & 5-year & 7-year \\
      \hline
      \multicolumn{5}{c}{Run 8 (varying $n_s^{\ell<100}, n_s^{\ell>100}$, $A_s$, and
      $R$)} \\
      \hline
      $n_s^{\ell<100}$ & 1.020 & 1.014 & 1.011 $_{0.949,\ 0.887}^{1.083,\ 1.149}$ &
      1.032 $_{0.961,\ 0.896}^{1.097,\ 1.162}$ \\
      $n_s^{\ell>100}$ & 0.985 & 0.993 & 1.006 $_{0.973,\ 0.949}^{1.037,\ 1.076}$ &
      1.012 $_{0.981,\ 0.955}^{1.051,\ 1.095}$ \\
      $\log(10^{10} A_s)$ & 3.151 & 3.140 & 3.120 $_{3.045,\ 2.959}^{3.176,\
      3.227}$ &
      3.103 $_{3.022,\ 2.930}^{3.164,\ 3.215}$ \\
      $R$ & 0.006 & 0.019 & 0.000 $_{0.000,\ 0.000}^{0.108,\ 0.228}$ & 0.015
      $_{0.000,\ 0.000}^{0.128,\ 0.257}$ \\
      \hline
      \multicolumn{5}{c}{Run 9 (varying $\tau$, $n_s^{\ell<100}$, $n_s^{\ell>100}$,
      $A_s$, and $R$)} \\
      \hline
      $\tau$ & 0.0888 & 0.0904 & 0.0905 $_{0.0733,\ 0.0573}^{0.1098,\ 0.1305}$
      & 0.0920 $_{0.0765,\ 0.0624}^{0.1088,\ 0.1264}$ \\
      $n_s^{\ell<100}$ & 1.018 & 1.026 & 1.035 $_{0.960,\ 0.891}^{1.100,\ 1.167}$ &
      1.041 $_{0.970,\ 0.900}^{1.111,\ 1.177}$ \\
      $n_s^{\ell>100}$ & 0.984 & 0.996 & 1.005 $_{0.976,\ 0.948}^{1.046,\ 1.092}$ &
      1.014 $_{0.985,\ 0.956}^{1.059,\ 1.103}$ \\
      $\log(10^{10} A_s)$ & 3.163 & 3.141 & 3.121 $_{3.048,\ 2.958}^{3.183,\
      3.235}$ & 3.105 $_{3.025,\ 2.934}^{3.167,\ 3.220}$ \\
      $R$ & 0.002 & 0.020 & 0.000 $_{0.000,\ 0.000}^{0.107,\ 0.237}$ & 0.015
      $_{0.000,\ 0.000}^{0.126,\ 0.257}$ \\
      \hline
      \multicolumn{5}{c}{Run 10 (varying $\tau$, $n_s^{\ell<100}$, $n_s^{\ell>100}$, and
      $A_s$)} \\
      \hline
      $\tau$ & 0.0927 & 0.0907 & 0.0949 $_{0.0752,\ 0.0576}^{0.1124,\ 0.1316}$
      & 0.0938 $_{0.0772,\ 0.0625}^{0.1092,\ 0.1275}$ \\
      $n_s^{\ell<100}$ & 1.032 & 1.042 & 1.040 $_{0.965,\ 0.897}^{1.104,\ 1.177}$ &
      1.043 $_{0.971,\ 0.907}^{1.111,\ 1.184}$ \\
      $n_s^{\ell>100}$ & 1.025 & 1.034 & 1.031 $_{1.001,\ 0.978}^{1.054,\ 1.081}$ &
      1.032 $_{1.011,\ 0.986}^{1.062,\ 1.089}$ \\
      $\log(10^{10}A_s)$ & 3.084 & 3.062 & 3.076 $_{3.035,\ 2.986}^{3.131,\
      3.173}$ & 3.061 $_{3.016,\ 2.971}^{3.112,\ 3.154}$ \\
      \hline
      \multicolumn{5}{c}{Run 11 (varying $\tau$, $A_s$, and
      $R$)} \\
      \hline
      $\tau$ & 0.0745 & 0.0867 & 0.0711 $_{0.0598,\ 0.0459}^{0.0885,\ 0.1023}$
      & 0.0824 $_{0.0708,\ 0.0589}^{0.0973,\ 0.1101}$ \\
      $\log(10^{10} A_s)$ & 3.246 & 3.198 & 3.244 $_{3.214,\ 3.186}^{3.272,\
      3.301}$ &
      3.192 $_{3.165,\ 3.141}^{3.218,\ 3.244}$ \\
      $R$ & $1\times 10^{-4}$ & $5\times 10^{-5}$ & 0.000 $_{0.000,\
      0.000}^{0.011,\ 0.028}$ & 0.000 $_{0.000,\ 0.000}^{0.021,\ 0.050}$ \\
      \hline
      \multicolumn{5}{c}{Run 12 (varying $n_s^{\ell<100}$, $n_s^{\ell>100}$, and $A_s$)} \\
      \hline
      $n_s^{\ell<100}$ & 1.023 & 1.049 & 1.001 $_{0.953,\ 0.889}^{1.083,\ 1.153}$ &
      0.929 $_{0.879,\ 0.827}^{0.980,\ 1.028}$ \\
      $n_s^{\ell>100}$ & 1.022 & 1.036 & 1.018 $_{0.999,\ 0.973}^{1.044,\ 1.065}$ &
      0.977 $_{0.967,\ 0.958}^{0.984,\ 0.992}$ \\
      $\log(10^{10} A_s)$ & 3.073 & 3.052 & 3.070 $_{3.028,\ 2.986}^{3.120,\
      3.171}$ & 3.162 $_{3.143,\ 3.120}^{3.186,\ 3.208}$ \\
      \hline
      \multicolumn{5}{c}{Run 13 (varying $A_s$ and $R$)} \\
      \hline
      $\log(10^{10} A_s)$ & 3.263 & 3.199 & 3.263 $_{3.254,\ 3.245}^{3.272,\
      3.280}$ &
      3.197 $_{3.189,\ 3.181}^{3.206,\ 3.214}$ \\
      $R$ & $4\times10^{-5}$ & $4\times10^{-5}$ & 0.000 $_{0.000,\ 0.000}^{0.011,\ 0.027}$ &
      0.000 $_{0.000,\ 0.000}^{0.020,\ 0.048}$ \\
      \hline
  \end{tabular}
  \caption{Unmarginalised ML points and
  one-dimensional peaks and MCI limits resulting from a likelihood analysis with an $\ell$-dependent $n_s$, using the best-fit values used by Zhao et al. Results from the 5-year and 7-year analyses are shown, and all runs were over the range
  $\ell=2-220$.
  MCI limits are given
  with both 68\% and 95\% confidence.}
  \label{tab:ssangml}
\end{table*}
\begin{table}
    \centering
  \renewcommand\arraystretch{1.2}
    \begin{tabular}{ccc}
        Parameter & ML value & One-dimensional peaks and MCI limits \\
        & & $X\ ^{68\%\uparrow,\ 95\%\uparrow}_{68\%\downarrow,\
        95\%\downarrow}$ \\
        \hline
        $\Omega_bh^2$ & 0.0227 & 0.0233 $_{0.0225,\ 0.0218}^{0.0241,\ 0.0250}$ \\
        $\Omega_{DM}h^2$ & 0.1096 & 0.1053 $_{0.0987,\ 0.0920}^{0.1119,\ 0.1184}$\\
        $\theta$ & 1.040 & 1.041 $_{1.038,\ 1.035}^{1.044,\ 1.047}$ \\
        $\tau$ & 0.0910 & 0.0928 $_{0.0753,\ 0.0606}^{0.1086,\ 0.1275}$\\
        $n_s^{\ell<100}$ & 1.010 & 1.011 $_{0.949,\ 0.880}^{1.097,\ 1.168}$ \\
        $n_s^{\ell>100}$ & 0.974 & 0.988 $_{0.966,\ 0.947}^{1.014,\ 1.043}$ \\
        $\log(10^{10}A_s)$ & 3.166 & 3.111 $_{3.032,\ 2.941}^{3.182,\ 3.246}$ \\
        $R$ & 0.011 & 0.009 $_{0.000,\ 0.000}^{0.081,\ 0.165}$\\
        \hline
    \end{tabular}
    \caption{ML points and marginalised one-dimensional peaks and MCI limits
    for a likelihood analysis for eight parameters (including a step-like spectral index) using the WMAP 7-year data. MCI limits are
    given with both 68\% and 95\% confidence.}
    \label{tab:ssfullan}
\end{table}
\begin{table}
  \centering
  \begin{tabular}{c|cc}
    \hline
    \hline
    Parameters varied & $\log\mathcal{L}$ (5-year) & $\log\mathcal{L}$
    (7-year) \\
    \hline
    $n_s$, $A_s$, and $R$ (original analysis) & 284.938 & 134.985 \\
    $n_s$ and $A_s$ (original analysis) & 284.244 & 134.622 \\
    $n_s^{\ell<100}$, $n_s^{\ell>100}$, $A_s$, and $R$ & 284.972 & 135.044 \\
    $n_s^{\ell<100}$, $n_s^{\ell>100}$, and $A_s$ & 284.243 & 134.622 \\
    $A_s$ and $R$ & 275.896 & 131.768 \\
    $\tau$, $n_s$, $A_s$, and $R$ (original analysis) & 284.981 & 135.115 \\
    $\tau$, $n_s$, and $A_s$ (original analysis) & 284.386 & 134.832 \\
    $\tau$, $n_s^{\ell<100}$, $n_s^{\ell>100}$, $A_s$, and $R$ & 285.027 & 135.166 \\
    $\tau$, $n_s^{\ell<100}$, $n_s^{\ell>100}$, and $A_s$ & 284.392 & 134.816 \\
    $\tau$, $A_s$, and $R$ & 276.034 & 131.739\\
    \hline
  \end{tabular}
  \caption[Comparison of ML points]{Log(likelihood) values of the ML points of various runs from both the
  first and second analyses using the best-fit values obtained in this paper. The $\ell$-range is always 2-220.}
  \label{tab:likesnml}
\end{table}
\begin{table}
  \centering
  \begin{tabular}{c|cc}
    \hline
    \hline
    Parameters varied & $\log\mathcal{L}$ (5-year) & $\log\mathcal{L}$
    (7-year) \\
    \hline
    $n_s$, $A_s$, and $R$ (original analysis) & 285.083 & 134.921 \\
    $n_s$ and $A_s$ (original analysis) & 284.385 & 134.517 \\
    $n_s^{\ell<100}$, $n_s^{\ell>100}$, $A_s$, and $R$ & 285.205 & 135.033 \\
    $n_s^{\ell<100}$, $n_s^{\ell>100}$, and $A_s$ & 284.384 & 134.490 \\
    $A_s$ and $R$ & 275.849 & 132.922 \\
    $\tau$, $n_s$, $A_s$, and $R$ (original analysis) & 284.113 & 134.929 \\
    $\tau$, $n_s$, and $A_s$ (original analysis) & 284.482 & 134.565 \\
    $\tau$, $n_s^{\ell<100}$, $n_s^{\ell>100}$, $A_s$, and $R$ & 285.274 & 135.090 \\
    $\tau$, $n_s^{\ell<100}$, $n_s^{\ell>100}$, and $A_s$ & 284.477 & 134.566 \\
    $\tau$, $A_s$, and $R$ & 276.077 & 132.923\\
    \hline
  \end{tabular}
  \caption[Comparison of ML points]{Log(likelihood) values of the ML points of various runs from both the
  first and second analyses, using the best-fit values used by Zhao et al. The $\ell$-range is always 2-220.}
  \label{tab:likesgml}
\end{table}
\subsubsection{Discussion}
From tables \ref{tab:ssannml} and \ref{tab:ssangml}, we see that at least two of our expectations have not been met, or only partially: We find no higher values for $R$ when using a step-like spectral index than when we assume the spectral index to be constant, and under no circumstances does the value of $R$ reach the value it gets in runs 1 and 5 from the previous analysis, or anything near it. Further, even though $n_s^{\ell<100}$ and $n_s^{\ell>100}$ behave somewhat as expected, in that $n_s^{\ell<100} > n_s^{\ell>100}$ (so that the spectrum is bluer for $\ell < 100$ than for $\ell > 100$), their distributions lie closer than the distributions for $n_s$ from runs 1 and 2 (compare run 8 with runs 1 and 2).

Comparing the likelihood values of the ML models, we see from tables \ref{tab:likesnml} and \ref{tab:likesgml} that our third expectation has not been met either: Including a step-like spectral index only marginally raises the maximum likelihood, which suggests that allowing for an $\ell$-dependency in the spectral index serves little purpose, as it means introducing a new parameter which does nothing to improve the fit of our  model to the data.

\subsection{Data analysis conclusions}
Our analysis seems to suggest that the claims concerning an $\ell$-dependent spectral index made in \cite{zhao2} and \cite{zhao3} does not hold up under closer scrutiny. It seems there is neither any reason to abandon the assumption of a constant spectral index, since the distributions do overlap for two different $\ell$ ranges and since splitting up the spectral index as above gives no better fit to data, nor that this assumption leads to a negligence of signs of gravitational waves in the CMB.

We saw that in some of the $2\leq\ell\leq100$ runs, we found a lower limit on $R$. However, we noted that this lower limit was almost zero, and that it was just a 1$\sigma$ lower limit anyway. Our analysis gave lower $R$ values than the analysis in the papers by Zhao et al., which probably is mostly due to our use of the exact likelihood functions from WMAP in contrast to the approximations made in those papers.

As we mentioned in our comments on the papers by Zhao et al., we do not agree that searches for gravitational waves should be limited to a smaller set of data. As long as our cosmological parametrisation is reasonable, the Bayesian framework will automatically ensure that any signs of gravitational waves will be detected. The analysis performed in this paper brings us one step closer to saying that our parametrisation indeed is reasonable.

In any case, we have seen that the only indications of deviations from the standard analyses are very weak, only on the level of one confidence interval, and any claims made based on such weak detections become little more than speculation. As mentioned in the papers by Zhao et al., ongoing experiments such as Planck \citep{planck} and QUIET \citep{quiet}, among others, should make us much better suited to answer such questions as those addressed in this paper.
\section{Theoretical considerations}
It seems plausible that at least some of the motivation for the papers \cite{zhao2} and \cite{zhao3} comes from the work of Grishchuk on relic gravitational waves. He discovered that in any accelerating universe, existing gravitational waves will be amplified as long as their wavelengths are long enough \citep{gris3}. It is also notable that Grishchuk, since the mid-90s, has been a voiced opponent of the inflationary paradigm, which also may have been a motivation for the data analyses described above.


In \cite{gris2}, density perturbations are evolved through three stages: an initial, possibly inflationary stage, a radiation dominated stage, and a matter dominated stage. The amplitude of the density perturbations are then compared to the amplitude of the gravitational waves in the matter dominated stage (which were derived in an earlier paper, see \cite{gris4}), and Grishchuk finds that the density perturbations do not dominate; instead, the gravitational waves are the dominant contribution to the CMB. This is contrary to the predictions of the standard derivation, in which density perturbations dominate.

In addition to the above papers, Grishchuk authored \cite{gris5} and \cite{gris6}, in which he presented other arguments against the inflationary paradigm.

The above papers were criticized in \cite{martin1}, to which Grishchuk responded with \cite{gris7}. The argument was ended with \cite{martin2}, though it seems to us to be a bit unclear as to what the final conclusions were. After these papers, Grishchuk turned to the question of initial conditions in \cite{gris8}, the derivation in which again gave a different result than the standard one. This treatment was criticized in \cite{lukash}.

Overall, there seems to be little agreement on who, exactly, are correct in their treatment. We will therefore go through some points of disagreement and try to offer our own thoughts on them.
\subsection{The continuity of $\mu$}
Working in the synchronous gauge with the metric
\begin{equation}
    g_{00} = -a^2, \ \ \ g_{0i} = 0, \ \ \ g_{ij} = a^2\left( (1+hQ\delta_{ij} + h_ln^{-2}Q_{,i,j}) \right),
    \label{eq:synchpert}
\end{equation}
where $n$ is the wavenumber, $h$ is the scalar perturbation, and $h_l$ is the longitudinal-longitudinal perturbation (so that $Q$ becomes the amplitude of the perturbations), Grishchuk aims to find solutions for $h$ and $h_l$, along with general variables that appear in the Einstein equations, for all the three stages described above. For these stages, the scale factor may be written as
\begin{equation}
    a = l_0|\eta|^{1+\beta}
    \label{eq:scalefac}
\end{equation}
where the value of $\beta$ depends on which stage we are considering, and $\eta$ is conformal time.

In addition, Grishchuk defines certain auxilliary variables:
\begin{equation}
    \alpha = \frac{\dot{a}}{a}, \ \ \ \gamma = 1-\frac{\dot{\alpha}}{\alpha^2},
    \label{eq:alphagammadef}
\end{equation}
where overdots signifies derivatives with respect to conformal time. $\gamma$ then becomes a 'smallness parameter' during inflation.

He then goes on to solve the Einstein equations for all variables in each of the three stages.
It turns out that all variables can be expressed in terms of $h$ and $h_l$ for a given stage. It is therefore sufficient to find solutions for these variables.

At the initial (possibly inflationary) stage driven by a scalar field, so that $\phi_0$ is the homogeneous part and the total field is written as
\begin{equation}
    \phi = \phi_0 + \phi_1Q,
    \label{eq:phitot}
\end{equation}
all variables are expressible through $h$, and so, Grishchuk needs only find a solution for this variable. To this end, he introduces the variable $\mu$, defined through
\begin{equation}
    u = \frac{a\sqrt{\gamma}}{a}\mu
    \label{eq:mudef}
\end{equation}
where $u = \dot{h} + \alpha\gamma h$. During the initial phase, $\mu$ becomes
\begin{equation}
    \mu = (n\eta)^{1/2}\left( A_1J_{\beta+\frac{1}{2}}(n\eta)+A_2J_{-(\beta+\frac{1}{2})}(n\eta) \right)
    \label{eq:musol}
\end{equation}
where $J$ denotes the spherical Bessel functions, $\beta$ is the variable introduced in eq. (\ref{eq:scalefac}), and $A_1$ and $A_2$ are constants to be determined by the initial conditions . It can be shown that $h$ can then be written as
\begin{equation}
    h = \frac{\alpha}{a}\int_{\eta_0}^{\eta}\mu\sqrt{\gamma}d\eta + \frac{\alpha}{a}C_i,
    \label{eq:hsol}
\end{equation}
where $C_i$ is a constant that arises because of the degrees of freedom that remain in the synchronous gauge, and will be fixed by a choice of coordinate system.

Grishchuk finds similar solutions for $h$ (and $h_l$) during the radiation-dominated and matter-dominated stages, with three new constants arising at each stage. Finally, he aims to join these solutions, claiming that $h$, $\dot{h}$, and $\dot{h}_l$ are continuous over the transitions, and that in addition, $\mu$ is continuous over the transition from the initial to the radiation dominated stage. Using this, he joins the three stages, and by choosing the coordinate system that moves with the matter fluid, he manages to express the unknown constants at the matter dominated stage in terms of the constants $A_1$ and $A_2$, which are to be determined by initial conditions. He further makes some approximations which reduce $\mu$ from eq. (\ref{eq:musol}) to 
\begin{equation}
    \mu(\eta_1) \approx \frac{A_1}{2^{\beta+\frac{1}{2}}\Gamma(\beta + \frac{3}{2})}(n\eta_1)^{\beta+1},
    \label{eq:muapprox}
\end{equation}
where $\Gamma$ denotes the gamma function, and $\eta_1$ is the time of the transition between the initial stage and the radiation dominated stage. This means that all unknown constants can be related to $A_1$.

Using this solution, Grishchuk works out the observable consequences of it, and comparing them to the corresponding results for gravitational waves, he finds that for the CMB quadrupole, the gravitational wave contribution should actually dominate over the contribution from the density perturbations. This does not fit well with the standard picture, in which density perturbations should be the dominant contributor.

In \cite{martin1}, it is pointed out that the variable $\mu$ can not be continuous across this transition, but rather, that the combination $\frac{\mu}{a\sqrt{\gamma}}$ is continuous. The authors then claim that assuming $\mu$ to be continuous is what leads Grishchuk to deviate from the standard result.

Then, in \cite{gris7}, Grishchuk agrees that $\mu$ indeed is discontinuous over the transition, but that this was a typo in the original paper, and that he never actually used this assumption in his derivation. Finally, in \cite{martin2}, the authors uphold their claim that this assumption was used.

Grishchuk never responded to this, and so it is hard to tell whether he finally agreed with Martin \& Schwarz or not. One may, however, verify who is right, by looking at eq. (\ref{eq:muapprox}) (as also pointed out by Martin and Schwarz). Here, Grishchuk approximates $\mu$ at the transition between the initial and radiation stages. However, as we have already mentioned, $\mu$ is discontinuous at the transition, which means that one must choose whether one wants to use the value of $\mu$ at the beginning of the transition, or at the end of it. Grishchuk has earlier set $\gamma=2$, which means that he intends to be at the end of the transition when he is doing the joining. However, eq. (\ref{eq:muapprox}) is an approximation of eq. (\ref{eq:musol}), which is valid during the initial phase only. Thus, eq. (\ref{eq:muapprox}) is the value of $\mu$ at the \emph{beginning} of the transition, not the end, as it should be. This means that Grishchuk did implicitly use the assumption of a continuous $\mu$, as otherwise, eq. (\ref{eq:muapprox}) would not be valid (since we are at the end of the transition).

Since we have
\begin{equation}
  \frac{\mu}{a\sqrt{\gamma}}(\eta_1-\sigma) \approx \frac{\mu}{a\sqrt{\gamma}}(\eta_1+\sigma)
  \label{eq:transcont}
\end{equation}
where $\sigma$ is some arbitrarily small time interval so that $\eta_1-\sigma$ is the beginning of the transition while $\eta_1 + \sigma$ is the end of the transition, we find that the value of $\mu$ at the end of the transition, which is what we are really after in eq. (\ref{eq:muapprox}), should be
\begin{equation}
  \mu(\eta_1+\sigma)\approx \mu(\eta_1-\sigma)\sqrt{\frac{2}{\gamma_i}}
  \label{eq:muend}
\end{equation}
where 2 is the value of $\gamma$ at the radiation stage, while $\gamma_i$ is its value at the initial stage (typically very small). Thus, eq. (\ref{eq:muapprox}) should really be multiplied by $\sqrt{\frac{2}{\gamma_i}}$. This introduces an amplifying factor which propagates through the rest of Grishchuk's derivation, again yielding the standard result.

Thus, it seems that at this point, Grishchuk's treatment does not hold up under closer scrutiny.
\subsection{Initial conditions}
Martin \& Schwarz adressed some of the other complaints Grishchuk voiced about the inflationary theory, namely, the usage of the constancy of the curvature perturbation $\zeta$ (Grishchuk claims that this variable is equal to zero, and thus it is meaningless to utilize its constancy in theoretical derivations) and the large amplification of the density perturbation $\Psi$ from the inflationary stage to the radiation dominated stage. Both these issues seems to us to be cleared up: Grishchuk seems to agree in \cite{gris7} that $\zeta$ is both constant and nonzero, and, though it seems he disagrees with the exact word use, he also agrees that there is an amplification of $\Psi$ from the inflationary stage to the radiation stage. However, even though $\Psi$ is amplified, it does not mean that it necessarily dominates over gravitational waves - in order to say anything about which component dominates, we must look at the initial conditions. 

This is exactly what Grishchuk does in \cite{gris8}. In this paper, he goes through the initial conditions of the early Universe both for gravitational waves and density perturbations, and claims that the standard way of quantizing the curvature perturbation $\zeta$ is incorrect, since (as he shows) the expectation value of the normalized perturbation, called $\bar{\zeta}$, is not $\hbar/2$ as it should be for a proper harmonic oscillator; it is proportional to the 'largeness factor' $1/\gamma_i$. This, Grishchuk says, is due to the fact that the quantum states that are used for quantization in the standard derivation are not ground states; they are so-called 'squeezed states'. By doing a Bogliubov transformation of these states, he ends up with new states that satisfies the ground state condition. With these initial conditions, he finds that the spectrum of $\zeta$ at horizon crossing is comparable to, not hugely amplified compared to, the spectrum of gravitational waves.

His treatment was criticized by Lukash in \cite{lukash}, in which the author says that Grishchuk makes an error in the normalization of $\bar{\zeta}$, and it therefore becomes an error to demand the expectation value of this variable to be that of the ground state of an harmonic oscillator. In short, Grishchuk blamed the states, when he should have blamed the normalization of $\zeta$.

Lukash makes some other points as well to back up his criticism, but this will do for our purposes. We do agree with Lukash that the normalization of $\zeta$ in Grishchuk's paper seems inconsistent with what he himself says: $\zeta$ is normalized in exactly the same way as $h$ (the tensor perturbation), namely:
\begin{equation}
    \bar{\zeta} = \frac{a_0 M_{\rm Pl}}{k}\zeta,
    \label{eq:zetanorm}
\end{equation}
where $M_{\rm Pl}$ is the Planck mass, and $a_0$ is the value of the scale factor at some initial time at which to fix the initial conditions (with $\gamma_0$ being the corresponding value of $\gamma_i$. This value changes very little during inflation, so we have $\gamma_0 \approx \gamma_i$).
However, he also says that when moving from the treatment of gravitational waves to density perturbations, one should always do the replacement $a\rightarrow a\sqrt{\gamma_i}$. This means that the normalized curvature perturbation really should contain an extra factor $\sqrt{\gamma_0}$, which gets rid of the factor $1/\gamma_i$ when calculating the expectation value of the $\bar{\zeta}$ spectrum.  Therefore it seems to us that Grishchuk's claims about the initial conditions have been rebutted by Lukash. 

\section{Conclusions}
In this paper, we have studied various claims concerning primordial gravitational waves: We first looked at the claims made in \cite{zhao2} and \cite{zhao3} concerning the detection of a nonzero contribution of gravitational waves in the WMAP 5-year and 7-year data, by limiting the multipole range to $\ell=100$. After first noting that limiting the multipoles thus is neither more correct (as claimed by Zhao et al.) nor optimal in these data analyses, we repeated their analyses. We found that by using CosmoMC and the official noise values instead of approximated values for the noise, the gravitational wave contribution lessened, and if lower limits for $R$ were found, they were typically small. We also found no reason to believe their claim that it is an error to assume a constant spectral index for all multipoles, as the distributions for $n_s$ overlapped for the $\ell=2-100$ and $\ell=101-220$ analyses.

Nevertheless, we tested this claim further by implementing an $\ell$-dependent spectral index, allowing $n_s$ to have two different values above and below $\ell=100$. Our findings were not what we expected them to be if the claims by Zhao et al. are correct: The values of $n_s^{\ell<100}$ and $n_s^{\ell>100}$ did not converge to their expected values, and even with such a step-like spectral index, the gravitational wave contribution did not rise to any significant level. 

We did all of the above analyses twice; once where the parameters that were not varied were given the same values as by Zhao et al., which were the mean values from the WMAP analyses, and once where we used values from our own initial analysis where we allowed for gravitational waves but used the relation $n_t = n_s-1$ in accord with the statements of Zhao et al. For both sets of analyses, the results were qualitatively the same, so the conclusions seem to be independent of which best-fit values one prefers to use.

Finally, we went through some of Grishchuk's earlier claims concerning primordial gravitational waves, as these were conceived to be at least part of the motivation behind the paper by Zhao et al. We reviewed the controversy that followed Grishchuk's claims, and offered our own thoughts on the matter. In short, it seemed to us as Grishchuk's claims were firmly rebutted and that he has not responded adequately to these rebuttals.

The search for gravitational waves in the CMB will undoubtedly continue, and satelites such as Planck will surely shed further light on this issue. Just as surely, there are bound to be surprises in the next generation of CMB data. However, based on the data we have today, the gravitational wave contribution to the CMB fits nicely with the Standard Model and the inflationary theory.



\end{document}